\documentclass[a4paper,aps,11pt,nofootinbib]{revtex4}

\usepackage{amsmath}
\usepackage{amssymb}
\usepackage{amsfonts}
\usepackage[dvipdfm]{graphicx}
\usepackage{color}
\usepackage{float}
\usepackage{comment}
\usepackage{ulem}
\usepackage{ascmac}
\usepackage{setspace}
\usepackage{graphicx}

\begin{document}

\def\N{{\cal N}^3}

\title{
Inhomogeneous Generalization of Einstein's Static Universe \\
 with Sasakian Space
}

\today

\begin{spacing}{1.0}
\hfill{OCU-PHYS 552}

\hfill{AP-GR 176}

\hfill{NITEP 125}
\end{spacing}

\author{Hideki Ishihara}
\email{ishihara@osaka-cu.ac.jp}
\author{Satsuki Matsuno}
\email{sa21s010@osaka-cu.ac.jp}
\affiliation{
 Department of Mathematics and Physics,
 Graduate School of Science, 
 Nambu Yoichiro Institute of Theoretical and Experimental Physics (NITEP),
 Osaka City University,
 Osaka 558-8585, Japan
}


\begin{abstract}
We construct exact static inhomogeneous solutions to Einstein's equations 
with counter flow of particle fluid and a positive cosmological constant 
by using the Sasaki metrics on three-dimensional spaces. 
The solutions, which admit an arbitrary function that denotes inhomogeneous number 
density of particles, are a generalization of Einstein's static universe. 
On some examples of explicit solutions, we discuss non-linear density contrast and deviation 
of the metric functions. 
\end{abstract}


\maketitle

\section{Introduction}

In the general theory of relativity, the investigation of solutions to 
Einstein's equations is an important task to understand the structure 
of the universe. 
Since it is hard to solve Einstein's equations, which are non-linear field equations 
with constraints, 
almost solutions are found under simplification using isometries. 

Among exact solutions with matter sources, one of the most important ones are 
cosmological solutions of the Friedmann-Lema\^itre-Robertson-Walker metric, 
which describe the homogeneous and isotropic universe. 
Less symmetric solutions are provided by the Lema\^itre-Tolman-Bondi solutions \cite{LTB}, 
where spherically symmetric dust fluid is a source of gravity. 
It is striking that the solutions admit arbitrary functions. 
The most known generalization of the Lema\^itre-Tolman-Bondi solutions are Szekeres's 
solutions \cite{Szekeres}, which admit no geometrical symmetry. 

In the solutions noted above, the matter sources are characterized by vanishing vorticity. 
In contrust, we propose exact solutions to Einstein's equations with 
a fluid of particles moving along geodesics with non-vanishing vorticity. 

The total spacetimes of the solutions are direct products of time and static 
three-dimensional space. 
We take the space homothetic to a three-dimensional Sasakian space \cite{Sasaki, Blair},  
and construct exact solutions with inhomogeneous fluid with vorticity. 
The solutions admit an arbitrary function that describes density of the fluid.

\section{Metric with Sasakian Space }

We consider a static metric 
\begin{align}
	ds^2 &= - dt^2 + ds_M^2, 
\label{metric}
\end{align}
where the metric of the three-dimensional space, $M$, is given by 
\begin{align}
	ds_M^2 &=  a^2 \left(d\theta^2 + h(\theta, \phi)^2~d\phi^2 \right)
			+ b^2 (d\psi+ f(\theta, \phi) ~d\phi)^2.
\label{metric_M}
\end{align}
In \eqref{metric_M}, $a, b$ are constants, $f(\theta, \phi)$ and $h(\theta, \phi)$ are functions 
to be determined later. 
The metric \eqref{metric} admits two unit Killing vectors 
\begin{align}
	\xi_{(t)} = \partial_t \quad\mbox{and}\quad 
	\xi_{(\psi)} = \frac{1}{b} \partial_\psi. 
\label{Killing}
\end{align}

The space $M$ is a fiber bundle: 
a one-dimensional fiber with the coordinate $\psi$ on 
a two-dimensional base space, $N$, with the coordinate $(\theta, \phi)$. 
We take 1-form basis as
\begin{align}
	\sigma^0:=dt, \quad
	\sigma^1:= a d\theta, \quad \sigma^2:=a h(\theta, \phi) d\phi, \quad
	\sigma^3:=b( d\psi+ f(\theta, \phi) ~d\phi), 
\label{basis}
\end{align}
so that the metric \eqref{metric} with \eqref{metric_M} is rewritten as
\begin{align}
	g_{ab}&= -\sigma^0_a\otimes\sigma^0_b +g^{M}_{ab}, 
\cr
	g^M_{ab} &= \sigma^1_a\otimes\sigma^1_b +\sigma^2_a\otimes\sigma^2_b +\sigma^3_a\otimes\sigma^3_b. 
\label{metric2}
\end{align}
Assuming the relation between the function $f$ and $h$ as 
\begin{align}
	h(\theta, \phi)=\partial_\theta f(\theta, \phi), 
\label{Sasaki_cond}
\end{align}
we have
\begin{align}
	d\sigma^3 = \frac{b}{a^2} \sigma^1 \wedge \sigma^2, 
\label{contact_form}
\end{align}
and $\sigma^3\wedge d\sigma^3 \neq 0$. 
The manifold $M$ that admits such a 1-form is called a contanct manifold, and it is known that 
the three-dimensional space $(M, g^M)$ in the form of \eqref{metric_M} 
with the condition \eqref{Sasaki_cond}, which admits the unit Killing vector, 
is homothetic to a three-dimensional Sasakian space. 
The equation \eqref{contact_form} means existence of vorticity of the vector field 
$\xi_{(\psi)}$, which is metric dual to $\sigma^3$. 

The Scalar curvature of two-dimensional base space $N$ is 
\begin{align}
	R_N=-\frac{2}{a^2}\frac{\partial^2_\theta h(\theta, \phi)}{h(\theta, \phi)}, 
\label{K_N}
\end{align}
and the Ricci curvature tensor of the total spacetime 
with respect to the basis \eqref{basis} is given by
\begin{align}
	&R_{ab}
		= \left( -\frac{b^2}{2a^4 } + \frac12 R_N \right) 
	(\sigma^1_a\otimes \sigma^1_b + \sigma^2_a\otimes \sigma^2_b)
	+\frac{b^2}{2 a^4} \sigma^3_a\otimes \sigma^3_b.
\label{Ricci_tensor}
\end{align}

\bigskip

\section{Counter flow fluid}

We consider a counter flow fluid consists of collision-less particles: 
one component, labeled with \lq$+$\rq, flows in the direction of 
$\xi_{(\psi)}$, and the other, labeled with \lq$-$\rq, flows oppositely. 
Namely, the 4-velocities, parametrized by the proper time, are given by 
\begin{align}
	u_{+}^a
	= \frac{1}{\sqrt{1-v^2}} \xi^a_{(t)} + \frac{v}{\sqrt{1-v^2}} \xi^a_{(\psi)} ,
\cr
	u_{-}^a
	= \frac{1}{\sqrt{1-v^2}} \xi^a_{(t)} - \frac{v}{\sqrt{1-v^2}} \xi^a_{(\psi)}, 
\label{u}
\end{align}
where $v$ is a function that depends only on $\theta$ and $\phi$. 
Each particle with $u_\pm$ obeys the geodesic equation, 
\begin{align}
	u_\pm^a\nabla _a u^b_\pm =0. 
\label{geodesic}
\end{align}
As for the congruence of the geodesics with the tangent vectors \eqref{u}, 
we see that the expansion vanishes, and the shear does not vanish if $v$ is not 
a constant. The vorticity that comes from $\xi_{(\psi)}$ is non-vanishing if $v\neq 0$. 

The number densities of particles of counter flow are assumed as $n_{+}=n_{-}= n/2$, 
where $n$ is a function on $N$. 
Then, the energy-momentum tensor of the partcle fluid is
\begin{align}
	T^{ab}
		&= \frac12 m n  ( u_{+}^a \otimes u_{+}^b + u_{-}^a\otimes u_{-}^b)
\cr
		&=  m n \left( \frac{1}{1-v^2} \xi_{(t)}^a \otimes \xi_{(t)}^b 
		+ \frac{v^2}{1-v^2} \xi_{(\psi)}^a \otimes \xi_{(\psi)}^b\right),
\label{Tab}
\end{align}
and 
\begin{align}
	{\rm tr~}T = -mn. 
\end{align}
The total angular momentum vanishes by the counter flow. 
Taking the limit $m\to 0$ and $v^2 \to 1$ with $m/(1-v^2)=\rm{finite}$,  
we can consider the energy-momentum tensor of null particles moving along the fiber. 
\bigskip

\section{Einstein's equation}

From \eqref{Ricci_tensor} and \eqref{Tab}, Einstein's equation with a cosmological constant, 
\begin{align}
	R_{ab} 
		&= T_{ab}-\frac12 ({\rm tr~}T)g_{ab}+\Lambda g_{ab}, 
\label{Ricci_eq}
\end{align}
yields 
\begin{align}
	&0 = 
	\frac12 m n(\theta, \phi) \left(\frac{1+v(\theta, \phi)^2}{1-v(\theta, \phi)^2}\right)
		-\Lambda,
\label{R00}
\\
	&-\frac{b^2}{2a^4}+\frac12 R_N
	= \frac12 mn(\theta, \phi) + \Lambda,
\label{R11}
\\
	&\frac{b^2}{2a^4} = \frac12 m n(\theta, \phi) 
	\left(\frac{1+v(\theta, \phi)^2}{1-v(\theta, \phi)^2}\right)+\Lambda.
\label{R33}
\end{align}
Here and hereafter, we set $8\pi G=1$. Taking a combination of \eqref{R00} and \eqref{R33}, 
we have 
\begin{align}
	\Lambda = \frac{b^2}{4a^4} >0, 
\label{positive_Lambda}
\end{align}
and from \eqref{R00} we see that the function $v(\theta, \phi)$ is 
expressed by the function $n(\theta, \phi)$ as
\begin{align}
	v^2(\theta, \phi) =  \frac{2\Lambda -m n(\theta, \phi)}{2\Lambda+m n(\theta, \phi)}. 
\label{v2}
\end{align}
Then, $mn(\theta, \phi)$ should be in the range
\begin{align}
	0 \leq mn(\theta, \phi) \leq 2\Lambda . 
\label{mn_range}
\end{align}
Under the relation \eqref{positive_Lambda} and \eqref{v2}, 
the four-dimensional Einstein equations reduce to the simple equation, 
\begin{align}
	R_N(\theta, \phi)= m n(\theta, \phi)+ 6\Lambda. 
\label{K_N_n}
\end{align}
We call it \lq reduced Einstein's equation\rq\ on the two-dimensional base space 
that means the scalar curvature of $N$ equal to the mass density of particles plus 
the cosmological constant. 

Since $R_N(\theta, \phi)$ is positive everywhere on $N$, 
we consider $N$, as far as it is simply connected, 
to be homeomorphic to the two-dimensional sphere, hereafter. 
We integrate \eqref{K_N_n} on $N$ as
\begin{align}
	\int_N R_N(\theta, \phi) dS = \int_N \left(m n(\theta, \phi)+6\Lambda\right) dS.
\label{int_K}
\end{align}
By using the Gauss-Bonnet theorem, the left-hand side of \eqref{int_K} 
is $8\pi$. 
Introducing average of number density by
\begin{align}
	\langle n\rangle:= A_N^{-1} \int_N n(\theta, \phi) dS, 
\end{align}
where $A_N$ denotes surface area of the base space $N$, 
we have
\begin{align}
	m \langle n\rangle + 6\Lambda  = 8\pi A_N^{-1}, 
\label{ave_n}
\end{align}
no matter how $n(\theta, \phi)$ is inhomogeneous.

The reduced Einstein's equation \eqref{K_N_n} with \eqref{K_N} is written in the form
\begin{align}
	&\partial_\theta^2 h(\theta, \phi)
	+ a^2 w(\theta, \phi)h(\theta, \phi)=0, 
\label{eq_h}
\\
	&w(\theta, \phi) := \frac12 mn(\theta, \phi) + 3\Lambda. 
\label{w}
\end{align}
We should note that \eqref{eq_h} is a linear ordinary differential equation with 
respect to $\theta$ for every fixed value of the coordinate $\phi$. 

We take $(\theta, \phi)$ to be the geodesic polar coordinate system, 
then the function $h(\theta, \phi)$ should satisfy 
\begin{align}
	h(0,\phi)=0, 
	\quad h(\theta,\phi+2\pi)=h(\theta,\phi), 
\label{BC_h}
\end{align}
and
\begin{align}
	\partial_\theta h(0,\phi)=1, 
\label{BC_h'}
\end{align}
in order to avoid the conical singularities at the pole $\theta=0$.

\bigskip

\section{Examples}

Here, we present simple examples of global solutions. 
We consider $(\theta, \phi)$ as a spherical coordinate on the base space $N$, 
homeomorphic to $S^2$, in the range $0\leq \theta\leq\pi,~ 0\leq \phi\leq 2\pi$, 
where $\theta=0, \pi$ correspond to the north and south poles. 
The function $h$ should satisfies
\begin{align}
 h(\theta,\phi+2\pi)=h(\theta,\phi), 
\label{BC_h_1}
\end{align}
and 
\begin{align}
	h(0,\phi)=h(\pi,\phi)=0, \quad
	\partial_\theta h(0,\phi)=-\partial_\theta h(\pi,\phi)=1, 
\label{BC_h_2}
\end{align}
so that the coordinate singularities at the both poles can be removed. 


\subsection{\sl Homogeneous cases: }

In the case that the number density of the particles, $n$, is constant, 
\eqref{K_N_n} means $R_N=const.$, i.e., the two-dimensional base space $N$ is 
a homogeneous $S^2$ with radius $a$, and $A_N=4\pi a^2$. 
Since $n=\langle n\rangle$, \eqref{ave_n} leads to 
\begin{align}
	w=\frac1{2}{mn+3\Lambda}=\frac1{a^2}, 
\label{a2}
\end{align}
then we have 
\begin{align}
	h =\sin\theta 
\end{align}
as the solution to \eqref{eq_h} with 
the boundary conditions \eqref{BC_h_1} and \eqref{BC_h_2},  
and the function $f$ is given by   
\begin{align}
	f=-\cos\theta. 
\end{align}
With the help of \eqref{positive_Lambda}, the metric becomes 
\begin{align}
	ds^2 &= - dt^2
		+ a^2 \left(d\theta^2 + \sin^2\theta~d\phi^2 
		+ 4a^2\Lambda (d\psi- \cos\theta~d\phi)^2\right). 
\end{align}
We assume the fiber is $S^1$, so that the three-dimensional space $M$ is 
a Hopf's fiber bundle\footnote{ Indeed, if $M$ is simply connected and complete, 
it is proved that the fiber is $S^1$ and $M$ is a Hopf's bundle \cite{Manzano}.
} that describes a squashed $S^3$. 
The \lq aspect ratio\rq\ of the radius of $S^1$ fiber to the radius of $S^2$ base space is given by 
\eqref{a2} as
\begin{align}
	\frac{b}{a}=2a \sqrt{\Lambda} = \sqrt{\frac{8\Lambda}{mn+6\Lambda}} 
		=\sqrt{1+\frac{v^2}{2+v^2}} \geq 1 . 
\end{align}
Namely, $M$ is a \lq prolate\rq\ three-dimensional sphere for nonvanishing $v$,  
where the metric admits five Killing vectors: 
$\xi_{(t)}, \xi_{(\psi)}$ and three on the base space $S^2$. 

In the null particles limit, i.e., $m\to 0$ and $v^2\to 1$, the aspect ratio takes 
the maximum value, $2/\sqrt{3}$. 
On the other hand, in the case that the particles at rest, i.e., $v=0$ and 
$mn=2\Lambda$, the aspect ratio becomes 1, and we have 
\begin{align}
	ds^2 &= - dt^2
	+ \frac{1}{4\Lambda} \left( d\theta^2 + \sin^2\theta~d\phi^2 
	+  ( d\psi- \cos\theta~d\phi)^2\right). 
\end{align}
This is the metric of Einstein's static universe, where the three-dimensional space 
is a round $S^3$. 
This spacetime admits seven Killing vectors: $\xi_{(t)}$ and six on $S^3$ including $\xi_{(\psi)}$. 

\bigskip

\subsection{\sl Axisymmetric cases: }

We consider the case that the system is inhomogeneous but 
symmetric under a rotation of $\phi$. 
Then, the functions $n$ and $h$ depend only on $\theta$. 
Then, \eqref{eq_h} reduces to the equation, 
\begin{align}
	&\frac{d^2 h(\theta)}{d\theta^2}  
		+ a^2 w(\theta) h(\theta) =0, 
\cr
	&w(\theta)=\frac12 m n(\theta) + 3\Lambda. 
\label{Strum_Liouville}
\end{align}
The boundary conditions of $h(\theta)$ are
\begin{align}
	&h(0)= h(\pi)=0, 
\label{BC}
\end{align}
and $h(\theta)$ should be nonvanishing in the region $0 < \theta <\pi$. 
The ordinary differential equation \eqref{Strum_Liouville} 
with the boundary conditions \eqref{BC} is a Strum-Liouville problem, 
where $a^2$ is the eigenvalue and $w(\theta)$ is the weight function. 

At the north and south poles, regularity of the geometry requires 
\begin{align}
	\partial_\theta h(0)=1, \quad \partial_\theta h(\pi)=-1, 
\label{regularity}
\end{align}
and the smoothness of the number density requires
\begin{align}
	\partial_\theta n(0)= \partial_\theta n(\pi)=0. 
\end{align}

As a special example, we consider 
\begin{align}
	n(\theta)=n_0 - n_1 \cos(2\theta), 
\end{align}
where \eqref{mn_range} requires that $n_0$ and $n_1$ are constants satisfying 
\begin{align}
	0 \leq n_0-|n_1|, \quad {\rm and}\quad  
	n_0 + |n_1| \leq 2\Lambda/m . 
\end{align}
In this case, \eqref{Strum_Liouville} reduces to the Mathieu equation in the form 
\begin{align}
	\frac{d^2 h}{d\theta^2} + \Big( p - 2 q \cos(2\theta) \Big) h =0, 
\label{Matheiu_eq}
\end{align}
where $p$ and $q$ are constant parameters given by
\begin{align}
	p:= \left(3\Lambda +  \frac12 m n_0\right)a^2, \quad {\rm and}\quad q:=\frac14 m n_1 a^2. 
\end{align}
The solutions without node that satisfy \eqref{BC} and \eqref{regularity} are 
\begin{align}
	h(\theta)=C~se_1(q, \theta), 
\end{align}
where $se_1(q, \theta)$ is the odd Mathieu function of order 1, 
and $C$ is the normalization constant given by
\begin{align}
	\frac{1}{C}=\left.\frac{d}{d\theta} se_1(q,\theta)\right|_{\theta=0}. 
\end{align}
The function $f(\theta)$ is a primitive function of $C se_1(q, \theta)$. 
The metrics composed of the functions $h(\theta)$ and $f(\theta)$ have three Killing vectors: 
$\xi_{(t)}$, $\xi_{(\psi)}$ and  $\partial_\phi$. 
For given $n_0$ and $n_1$, the parameter $a$ is determined 
so that $p$ should be the characteristic value of $se_1(q, \theta)$, 
then $h(\theta)$ satisfies \eqref{BC} and \eqref{regularity}. 

We consider the case that the mass density varies maximally in \eqref{mn_range}, 
namely,
$mn_{min}=0$ and $mn_{max}=2\Lambda$.
Setting $ n_0=|n_1|=\Lambda/m $, we have two cases 
for $(i)~ q=\frac{1}{4}\Lambda a^2$ and $(ii) ~~q=-\frac{1}{4}\Lambda a^2$:  
\begin{align}
	(i)\quad n(\theta)=\Lambda(1-\cos2\theta)\quad: \quad 
	\mbox{sparse at the poles and dense at the equator}, 
\cr
	(ii)\quad n(\theta)=\Lambda(1+\cos2\theta)\quad: \quad
	\mbox{dense at the poles and sparse at the equator}. \nonumber
\end{align}
In these cases, 
$p=\frac{7}{2}\Lambda a^2$ should be the the characteristic value 
of the Mathieu functions $se_1\left(\pm\frac{1}{4}\Lambda a^2,\theta\right)$, 
then $a$ and related quantities are determined numerically as 
\begin{align}
	(i)\quad&a= 0.5162 \Lambda^{-1/2}, \quad A_N= 1.09003\Lambda^{-1} \pi, 
		\quad m\langle n\rangle= 1.33922\Lambda,
\cr
	(ii)\quad&a= 0.5545 \Lambda^{-1/2}, \quad A_N= 1.19876\Lambda^{-1} \pi, 
		\quad m\langle n\rangle= 0.673552 \Lambda . \nonumber
\end{align}
As a reference, 
$a=(1/2)\Lambda^{-1/2}, ~A_N=\Lambda^{-1} \pi$,  and $m\langle n\rangle=mn=2\Lambda$ 
for Einstein's static universe. 
While the geometrical quantities take the similar values in these cases, i.e., 
$a\sim 0.5\Lambda^{-1/2}$, $A_N \sim \Lambda^{-1}$,  
the avaraged mass density of the case $(i)$ takes almost double of the case $(ii)$. 

\bigskip

\subsection{\sl Non-axisymmetric cases: }
\label{nonAxis}

On the assumption of the metric \eqref{metric_M} with the boundary conditions \eqref{BC} 
and \eqref{regularity}, the $\phi=const.$ curves, which connect the north pole 
and the south pole, are geodesics on the base space $N$, 
and all these curves have the same length, $\pi a$.   
Then, the inhomogeneous global solutions obtained in this paper are such class of 
special solutions. 

Although it is possible, in principle, to solve the equation \eqref{eq_h} with \eqref{w} 
for a given smooth function $n(\theta, \phi)$, 
it is hard to represent the solutions by using well-known special functions. 
Starting from $f(\theta, \phi)$, however, 
we can easily present a set of functions $h(\theta, \phi)$ and $n(\theta, \phi)$ 
expressed by combinations of the trigonometric functions as exact solutions. 

As an exact solution, we present metric functions 
\begin{align}
	f(\theta,\phi)&=-\cos\theta +\beta\sin^5\theta\cos\phi, 
\label{inhommo_metric_f}
\\
	h(\theta,\phi)&=\sin\theta+5\beta\sin^4\theta\cos\theta\cos\phi, 
\label{inhommo_metric_h}
\end{align}
and the mass density function 
\begin{align}
	mn(\theta,\phi)	&= -6\Lambda 
	+\frac1{a^2} \left(2
		- ~\frac{30\beta\sin4\theta\cos\phi }
			{1+5\beta\sin^3\theta\cos\theta\cos\phi } \right), 
\label{density_n}
\end{align}
where $\beta$ is a positive parameter that denotes an amplitude of inhomogeneity.
The metrics composed of the functions \eqref{inhommo_metric_f} and 
\eqref{inhommo_metric_h} have only two Killing vectors, $\xi_{(t)}$ and $\xi_{(\psi)}$, 
if $\beta\neq 0$, 
while in the special case $\beta=0$, the solutions reduce to the homogeneous cases 
discussed above. 

For the functions \eqref{inhommo_metric_f}, \eqref{inhommo_metric_h} 
and \eqref{density_n}, which have inhomogenity, 
the surface area $A_N$ and averaged mass density $m\langle n\rangle$ are obtained as
\begin{align}
	A_N 
	=4\pi a^2 
\quad\mbox{and}\quad 
	m\langle n\rangle 
	=\frac{2}{a^2} - 6 \Lambda.
\end{align}
These quantities, independent of the parameter $\beta$ explicitely, 
are the same forms in the homogeneous case. 

\begin{figure}[!h]
\centering
\includegraphics[width=12cm]{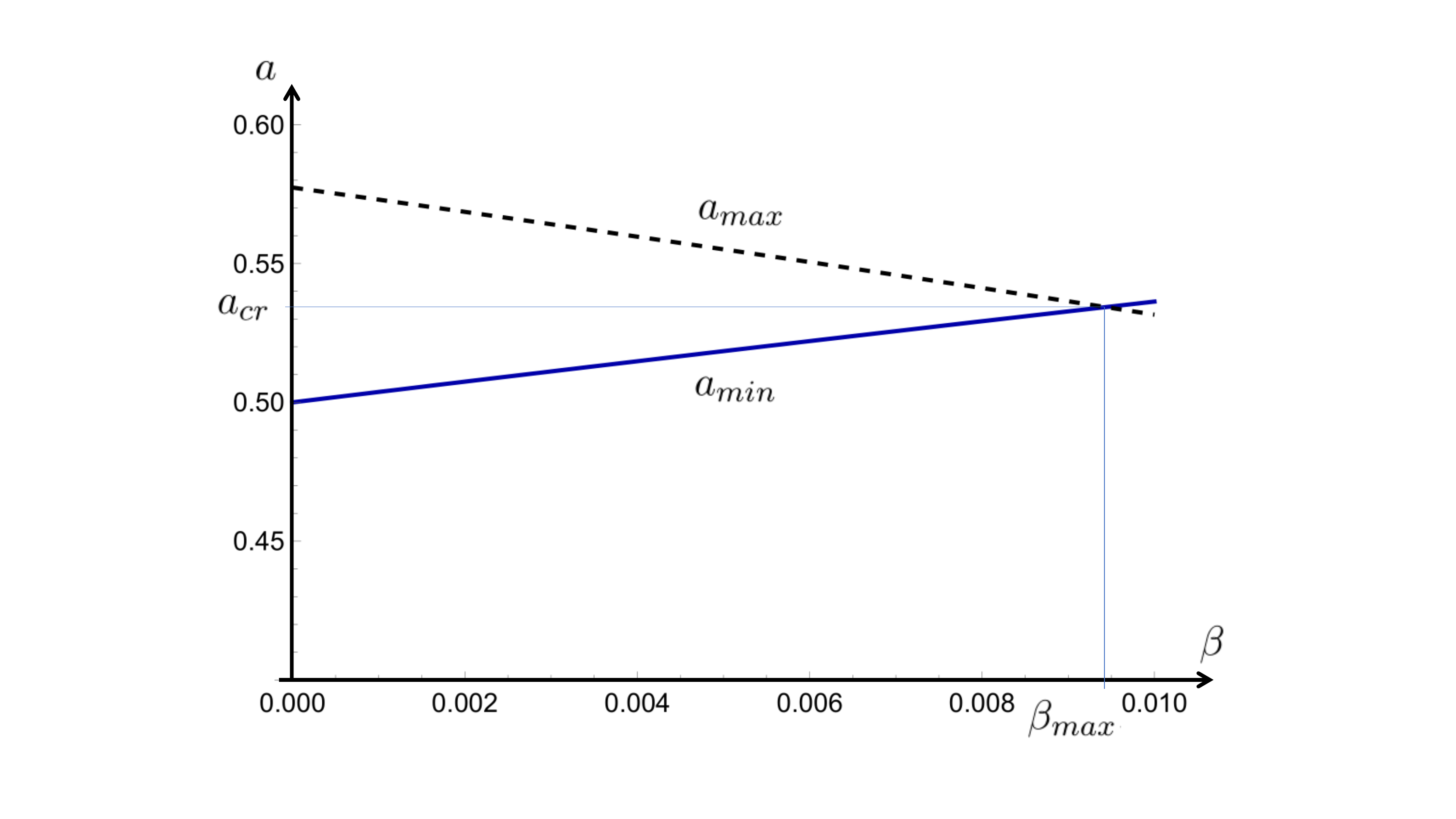}
\caption{
The upper bound, $a_{max}$, and lower bound, $a_{min}$, of $a$ are depicted as 
functions of $\beta$. At $\beta=\beta_{max}$, $a_{max}$ and $a_{min}$ coincide 
with $a_{cr}$. 
\label{a_vs_beta}
}
\end{figure}

The parameter $\beta$ and $a$ are limited as $0\leq \beta\leq\beta_{max}$ and 
$a_{min}\leq a \leq a_{max}$ 
so that $mn(\theta, \phi)$ satisfies \eqref{mn_range}.   
In the case that $mn(\theta,\phi)$ varies maximally, 
namely it takes $0$ and $2\Lambda$ elsewhere, 
$\beta$ becomes upper bound $\beta_{max}$, and $a_{min}$ and $a_{max}$ coincide 
with a value $a_{cr}$. 
In Fig.\ref{a_vs_beta}, $a_{min}$ and $a_{max}$ are shown as functions of $\beta$,  
where $\beta_{max}\sim 0.009436$ and $a_{cr}\sim 0.5343$, numerically\footnote{ 
A rough estimation of $\beta_{max}$ and $a_{cr}$ is given in Appendix. 
}.
It is interesting that even for the non-linear density contrast, 
$(n_{max}-n_{min})/(n_{max}+n_{min})=1$, 
\eqref{inhommo_metric_h} and \eqref{inhommo_metric_f} with $\beta=\beta_{cr}$ means 
that the deviation of the metric functions 
is small in the order of $1/100$.

\bigskip

\section{Summary}

We have constructed exact static inhomogeneous solutions to Einstein's equations 
with counter flow of particle fluid and a positive cosmological constant. 
The three-dimensional space of the solution is homothetic to a Sasakian space 
that consists of $S^1$ fibers on a $S^2$ base space. 
The solutions admit two unit Killing vector fields: 
timelike Killing vector field of the static spacetime, 
and spacelike Killing vector field that is tangent to the fiber. 
The unit Killing vector tangent to the fiber, whici is metric dual to the contact 
form of the three-dimensional space, is geodesic tangent and has non-vanishing rotation. 
Particles of the fluid move along geodesics whose tangent vectors are linear 
combinations of the two Killing vectors metioned above. 
Then, the geodesic congruences of the particles have non-vanishing vorticity. 

On these assumptions, we have obtained reduced Einstein's equations 
on the two-dimensional base space that makes a relation of the scalar curvature  
with the mass density of the particles, and the cosmological constant. 
The equation has a form of linear differential equation for the metric function. 
We have found exact solutions to the differential equation, where 
the number density of particles has non-linear inhomogeneity denoted by an arbitrary 
function on the base space. 
The solutions are inhomogeneous generalizations of Einstein's static universe.  
We have presented examples of exact solutions explicitely, 
and we observed that the deviation of the metric is small in the order of $1/100$ 
for non-linear density contrast of the particles.  

As is well known that Einstein's static universe is dynamically unstable. 
Similarly, the solutions obtained in this paper would be unstable. 
It is interesting problem to extend the solutions to expanding ones with inhomogeneity.

\section*{Acknowledgements}

We would like to thank K.-i. Nakao, H. Yoshino, H. Itoyama, and J. Inoguchi 
for valuable discussion. 

\appendix
\section{Rough estimation of $\beta_{max}$ and $a_{cr}$ of the model in Section \ref{nonAxis}}
The upper and lower limits of $a$ are determined by $mn=0$ and $mn=2\Lambda$, 
respectively. Then we have 
\begin{align}
	&a_{max}^2 =\frac1{3\Lambda } w(\theta_-,\phi_-;\beta),
\cr
	&a_{min}^2=\frac1{4\Lambda } w(\theta_+,\phi_+;\beta),
\label{a_minmax}
\end{align}
where
\begin{align}
	&w(\theta,\phi;\beta):=
		1 - ~\frac{15\beta\sin4\theta\cos\phi }
			{1+5\beta\sin^3\theta\cos\theta\cos\phi } , 
\label{w_beta}
\end{align}
and $(\theta_+,\phi_+)$ and $(\theta_-,\phi_-)$ give the maximum and mminimum of 
$w(\theta,\phi;\beta)$, respectively.  
We see that 
\begin{align}
	\partial_\phi w(\theta,\phi;\beta) =0  \quad \mbox{for } \phi=0,~\pi, 
\end{align}
and $w(\theta,\phi;\beta)$ is invariant under 
\begin{align}
	\phi \to \phi+\pi, \quad \theta \to \pi -\theta, 
\end{align}
then we fix $\phi=0$. 
The parameter $\beta$ should be small for positive $w(\theta,\phi;\beta)$, 
then the minimum of $w(\theta,\phi;\beta)$ is attained for
\begin{align}
	\sin 4\theta \sim 1, \quad \sin^3\theta\cos\theta<0,
\end{align}
and maximum for
\begin{align}
	\sin 4\theta \sim -1,\quad \sin^3\theta\cos\theta<0. 
\end{align}
Then, we have approximately 
\begin{align}
	(\theta_+,\phi_+) \sim  \left(\frac78 \pi,0\right) \quad \mbox{and} \quad 
	(\theta_-,\phi_-) \sim  \left(\frac58 \pi,0\right) .
\label{theta_pm}
\end{align}
We expand \eqref{a_minmax} by the small parameter $\beta$ upto the second order as
\begin{align}
	a_{max}^2\sim\frac{1}{3\Lambda}(1-15\beta\sin(4\theta_-)
		+75\beta^2 \sin(4\theta_-)\sin^3\theta_-\cos\theta_-),
\cr
	a_{min}^2\sim\frac{1}{4\Lambda}(1-15\beta\sin(4\theta_+)
		+75\beta^2 \sin(4\theta_+)\sin^3\theta_+\cos\theta_+). 
\end{align}
For small $\beta$, $a_{max}$ and $a_{min}$ are almost linear functions of $\beta$ 
as is seen in Fig. \ref{a_vs_beta}. 
By equating $a_{max}^2$ to $a_{min}^2$, and using \eqref{theta_pm}, we can estimate 
$\beta_{max}\sim 0.00944$, and $a_{cr}=a_{min}=a_{max}\sim 0.534$.


\end{document}